\documentclass[aps,prb,twocolumn,showpacs,a4paper,unsortedaddress,amsmath,amssymb,byrevtex]{revtex4}

\usepackage[latin1]{inputenc}
\usepackage{graphicx}
\usepackage{dcolumn}
\usepackage{bm}
\usepackage{times}

\begin{document}
\preprint{ffuov/05-04}

\title{Single channel conductance of H$_2$ molecules attached to platinum or palladium electrodes}

\author{V. M. Garc\'{\i}a-Su\'arez$^1$\footnote{Present address: Department of
Physics, Lancaster University, Lancaster, LA1 4YB, U. K.}}
\author{A. R. Rocha$^2$}
\author{S. W. Bailey$^3$}
\author{C. J. Lambert$^3$}
\author{S. Sanvito$^2$}
\author{J. Ferrer$^1$}
\affiliation{$^1$ Departamento de F\'{\i}sica, Universidad de Oviedo, 33007 Oviedo, Spain}
\affiliation{$^2$ Physics Department, Trinity College, Dublin 2, Ireland}
\affiliation{$^3$ Department of Physics, Lancaster University, Lancaster, LA1 4YB, U. K.}

%\homepage{condmat.uniovi.es/victor}

\date{\today}

\begin{abstract}
We report a detailed theoretical study of the bonding and
conduction properties of an Hydrogen molecule joining either platinum or
palladium electrodes. We show that an atomic arrangement where the
molecule is placed perpendicular to the electrodes is unstable for all
distances between electrodes. In contrast, the configuration where the
molecule bridges the electrodes is stable in a wide range of distances.
In this last case the bonding state of the molecule does
not hybridize with the leads and remains localized within the
junction. As a result, this state does not transmit charge so that
electronic transport is carried only through the anti-bonding state.
This fact leads to conductances of 1 $G_0$ at most, where $G_0=2e^2/h$.
We indeed find that G is equal to 0.9 and 0.6 $G_0$ for Pt and Pd contacts
respectively.
\end{abstract}


\maketitle

The main obstacle to using single molecules as conducting elements
\cite{Avi74} is the difficulty in establishing stable and
reproducible contacts to metallic electrodes. Conductance
quantization in atomic constrictions was predicted several years
ago \cite{Fer88}, but the practical realization in single-molecule
contacts only became possible in the mid-90's, with the advent of
the scanning tunneling microscope \cite{Dor95,Bum96,And96} and
mechanically controllable break junctions \cite{Ree97}.

An important testing ground for both theory and experiment is the
simplest of all molecular bridges, namely the H$_{2}$ molecule.
This molecule has just one bonding and one anti-bonding state in
the available energy range. Therefore it should be possible to
predict the effect  of tuning a range of adjustable parameters,
including the position and orientation of the molecule, the
distance and voltage between the electrodes and the materials that
make up the electrodes. Despite the apparent simplicity of this
junction, there is currently no agreement about what are the
position and orientation of the molecule for given separation
between the electrodes, even for zero applied bias.

Smit and coworkers \cite{Smi02} found that the conductance histograms of the
molecule, sandwiched between Pt leads, had a sharp peak at about 0.9 $G_0$
($G_0=2e^2/h$ is the conductance quantum) and therefore argued
that electronic transport was dominated by a
single channel with an almost perfect transmission. In addition, from the experimental
phonon spectra they concluded that the molecule bonds to the
electrodes in a bridge configuration (BC), i.e. with the H$_2$ bond axis
parallel to the transport direction (see Fig.1(a)). This was also confirmed by
Cuevas et al. \cite{Cue03}.

A different interpretation was given by Garc\'{\i}a and co-workers
\cite{Gar04} who showed theoretically that an arrangement where
the molecule bonds perpendicularly to the transport direction
(perpendicular configuration, PC) has a lower energy, and
therefore should also be the preferred atomic configuration (see
Fig. 1(b)). Their simulations gave conductances of order 1~$G_0$
and 0.2~$G_0$ for PC and BC, respectively. Very recently, Thygesen
and collaborators \cite{Thy04} challenged Garc\'{\i}a's results.
They performed a careful study of the vibrational spectra
providing further theoretical evidence in favor of the initial
interpretation of Smit et al. \cite{Smi02}. They found that the
current is carried by the anti-bonding state of the molecule with
a conductance of the order of 1~$G_0$, but in agreement with
Garc\'{\i}a, they predicted that the PC has a lower energy than
the BC, when the separation between leads is small.

As a summary, the analysis of the experimental data performed in
terms of vibration modes and conduction channels indicates that
the H$_2$ molecule places itself in a BC configuration and
therefore, current is carried by a single channel
\cite{Smi02,Cue03}, identified to be the anti-bonding state of the
molecule \cite{Thy04}. But the energetic analyses carried out so
far indicate that the PC arrangement is more stable and should
therefore be realized in a wide range of separation distances
between the electrodes. Since the transport in this arrangement is
carried by two channels, the whole interpretation of data in terms
on a single channel seems jeopardized. Even if the PC
configuration turns out not to be energetically favourable, there
remains the question of why the bonding state does not carry any
current and how does this fact manifest in the electronic spectra
of the molecular setup. To resolve these contradictions and open
questions, we present a detailed investigation of the stability,
electronic structure and conductance of the H$_2$ molecule
attached to either Pt or Pd electrodes.

First, we have simulated, in adition to the atomic configurations
analysed previously by other authors \cite{Gar04,Cue03,Thy04}, a
number of new arrangements that could very likely occur, and that
we depict in Figs. 1 and 2, respectively. We show that, for short
distances, H$_2$ molecules prefer to bind rather to the surface of
one of the electrodes than in between them. Therefore, at those
distances, the atomic constriction involves only the Pt electrodes
and current is carried directly through Pt atoms, which involves
several channels. PC begins to compete in energy with these atomic
configurations in a narrow range of intermediate distances. Since
all such configurations are likely to occur, and each of them has
a different number of transport channels, we do not expect a peak
in conductances histograms. As the distance increases further, the
energy of BC starts to be competitive and eventually becomes the
most stable configuration since for long enough distances all the
other configurations break.

\begin{figure}
\includegraphics[width=\columnwidth,height=8cm]{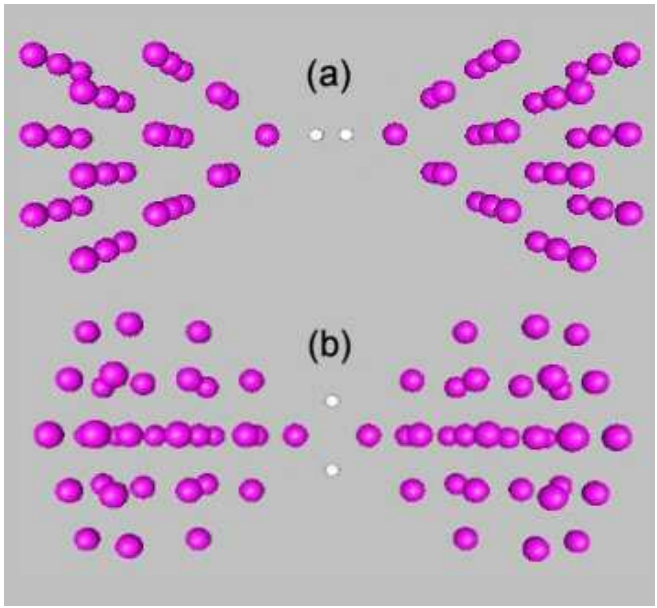}
\caption{\label{Fig1} Atomic configuration of the scattering
region for the BC (a) and PC (b) arrangements. Shaded and white
balls indicate Pt (or Pd) and H atoms, respectively. Fig. (b) has
been rotated 45 degrees around the z-axis with respect to Fig.
(a), to achieve a better view of the orientation of the Hydrogen
molecule. }
\end{figure}

Second, we have studied the Density of States (DOS) projected on
the different atoms of the junction for both PC and BC
arrangements. For the BC configuration, we find that the bonding
state of the molecule does not hybridize with the Pt electrodes
and remains a localized state of the junction.
This state is therefore a closed conduction channel that can
not swap charge from one electrode to the other. On the
contrary, the anti-bonding state completely spreads out due to
hybridization, and therefore becomes the single conduction channel
of the junction.
For the PC configuration, on the contrary, both bonding and
anti-bonding states hybridize strongly to the Pt electrodes. Therefore,
in this atomic arrangement, both states become conduction channels
that exchange electrons between the electrodes.

Third, a detailed study of the transmission coefficients shows
that T(E)  is essentially constant and equal to 0.9-1.0 for a wide
range of energies around E$_F$ and distances about the equilibrium
distance of the BC configuration, as expected for a junction with
a single conduction channel. This shows that the anti-bonding
channel has almost perfect transmission. The transmission
coefficient T(E) shows strong variations for the PC arrangement.
It indeed varies from about 1 to 2.5 in a relatively narrow range
of energies. This fact reveals that at least three channels are
open in this configuration.

Finally, we have also studied another closely related experiment,
performed by Csonka et al \cite{Cso04}, where H$_{2}$ was placed
between Pd leads. In this case, the conductance histogram showed a
single peak centered at 0.6 $G_0$, provided the gas pressure
inside the chamber was high enough. Our simulations show that
transport in BC is also carried only through
the anti-bonding state. We find that the
total conductance is smaller for Pd than for Pt in both arrangements.
The value we obtain for the BC configuration agrees with the
experimental result, and indicates that the BC configuration is
also realized for these junctions.

\begin{figure}
\includegraphics[width=\columnwidth,height=11cm]{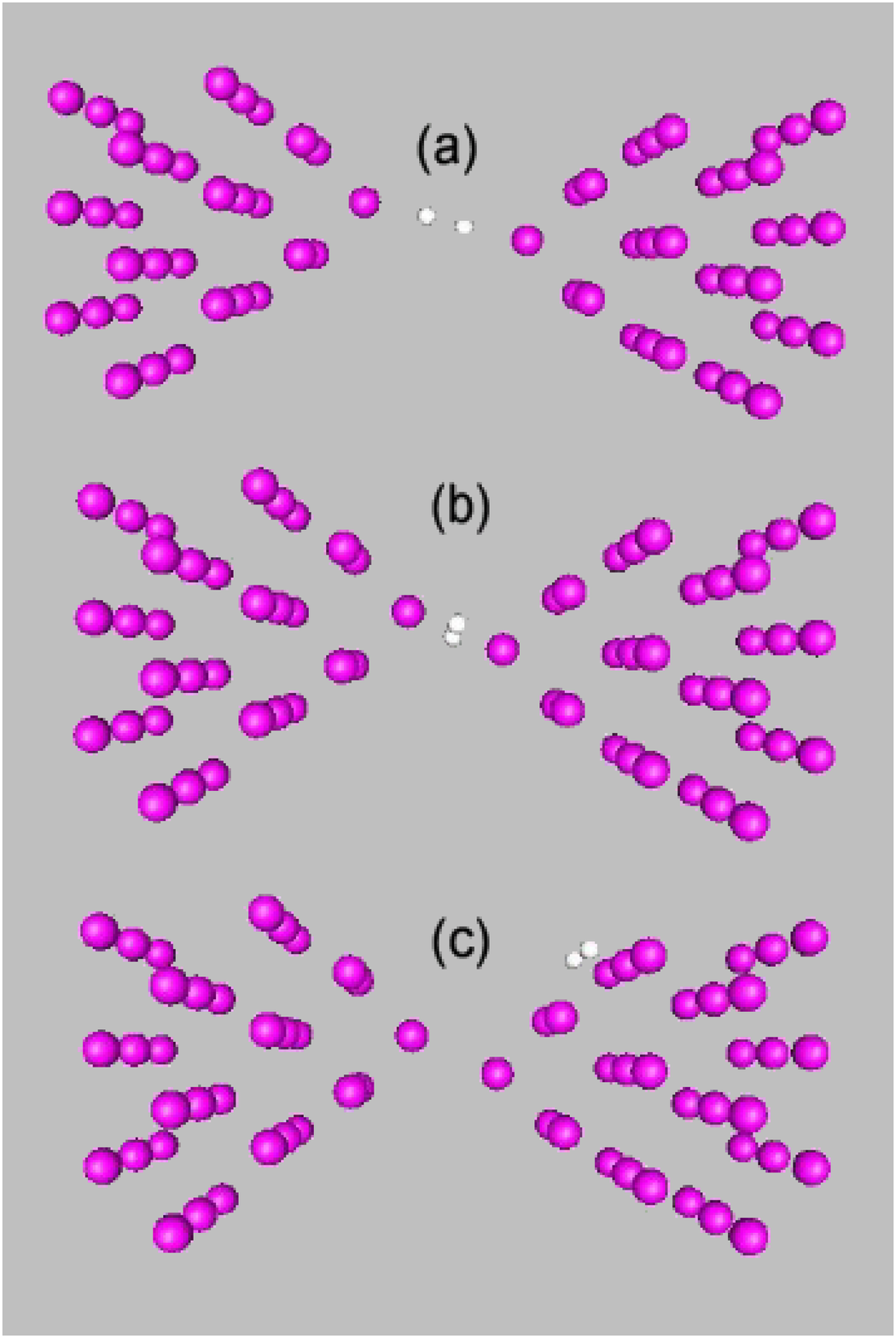}
\caption{\label{Fig2} New atomic configurations studied in this
work: BC zigzag (a), PC zigzag (b) (top view of the molecule)
and zigzag with the molecule chemisorbed on the surface (CZ) (c).}
\end{figure}

We use our recently developed {\it ab initio} quantum transport
code SMEAGOL~\cite{Smeagol,Natmat} for determining both the stable and
metastable H$_2$/electrode atomic configurations and for computing
their low-bias conductances. SMEAGOL combines the non-equilibrium
Green's function (NEGF)
formalism~\cite{Kel65,Car71,Fer88,Dat95,Pal01,Roc04} with density
functional theory (DFT)~\cite{Sha65}, using the numerical
implementation contained in the SIESTA code \cite{Sol02,Basis}.
The system (molecule plus leads) is ideally divided in three
regions: namely a left and a right lead, and a scattering region
comprising the H$_2$ molecule and part of the leads themselves \cite{Fis82}.
The two leads act as current/voltage probes and are assumed to be
in equilibrium, with well-defined chemical potentials. In
contrast, the potential of the scattering region is calculated
self-consistently for each applied bias \cite{Kel65,Car71,Dat95}.
For this purpose we define the Green's function of the scattering
region in the presence of the leads as
\begin{equation}
\hat{G}= \lim_{\delta \to 0}
\left[(E+i\delta)\hat{S}-\hat{H}_S[\rho]-\hat{\Sigma}_{L}-\hat{\Sigma}_{R}\right]^{-1}\:,
\label{GF}
\end{equation}
\begin{figure}
\includegraphics[width=7cm,height=7cm]{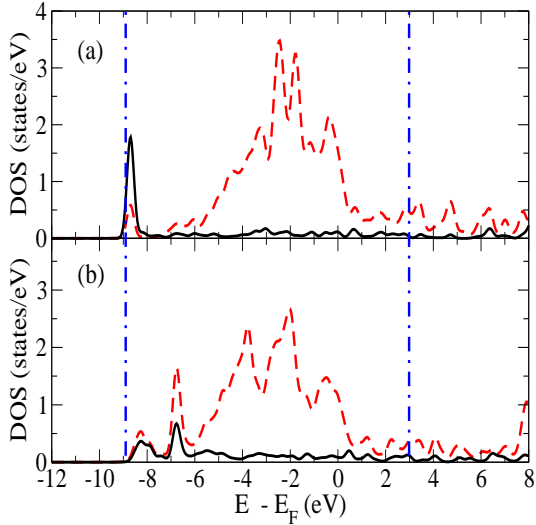}
\caption{\label{Fig3} Pt electrodes: DOS projected on one of the
Hydrogen atoms and the Pt atom at one of the two apexes (solid and
dashed lines, respectively) in (a) BC and (b) PC configurations.
Dashed-dotted vertical lines: position of the bonding and
anti-bonding energy levels of an isolated H$_2$ molecule. Energies
have been referred to the Fermi energy of the Pt leads.}
\end{figure}

\noindent where $\hat{H}_S[\rho]$ is DFT Hamiltonian and
$\hat{\Sigma}_{L}$ and $\hat{\Sigma}_{R}$ are the self-energies
respectively for the left and right lead. This allows us to
evaluate the density matrix
\begin{equation}
\rho=\int \frac{dE}{2\pi}
\hat{G}\left[\hat{\Gamma}_{L}f(E-\mu_{L})+
\hat{\Gamma}_{R}f(E-\mu_{R})\right]\hat{G^{\dagger}}\;,
\label{DM}
\end{equation}
with
$\hat{\Gamma}_{\alpha}=i[\hat{\Sigma}_{\alpha}-\hat{\Sigma}_{\alpha}^\dagger]$.
Since the DFT Hamiltonian $H_S$ depends solely on the density
matrix, equations (\ref{GF}) and (\ref{DM}) can be iterated until
reaching self-consistency. Then the current is extracted using a
Landauer-type formula \cite{Roc04} $G(E)=G_0\,T(E)$, where $T(E)$
is the transmission coefficient for electrons of energy $E$
relative to the Fermi energy $E_F$.

In what follows, the leads have an fcc crystalline structure and
are oriented along the (001) direction. Their unit cell is a slab
containing two 3x3 atomic layers. The scattering region, presented
in Figs.~1 and 2, consists of an H$_{2}$ molecule attached to the
tips of two pyramids of Pt or Pd atoms. These pyramids can be
viewed as a sequence of two atomic fcc (001) layers  comprising 4
and 1 atoms respectively. The scattering region also includes
three bulk Pt (or Pd) buffer layers with the same atomic
configuration of the leads, that are seamlessly attached to the
pyramids. The Hydrogen molecule is placed both in the BC and in
the PC setup, as shown in Figs.~1 (a) and (b) respectively, or
elsewhere. For the simulations, periodic boundary conditions are
applied in the basal plane, and 4 irreducible $k$-points in the
2-dimensional Brillouin zone are used \cite{k-points}. We label
every simulation according to the distance $d$ between the
innermost buffer layers of the scattering region. For a given
distance $d$, we always relax the pyramids and the Hydrogen atoms,
and compute the total energy at zero applied bias.

Consider first the case of Pt electrodes. We find that the
molecule in isolation has an equilibrium distance of 0.81 \AA, and
a bonding-antibonding energy gap of about 11.5 eV. This bond is
weakened when the molecule is attached to the electrodes. The
relaxed distances are 0.99 \AA\, for BC and 2.46 \AA\, for PC,
which indicate that this last configuration is close to a
dissociated state. The Densities of States projected onto either Hydrogen
or the Pt atom at the apex are shown in Fig. 3 for both BC and PC
configurations. Fig 3 (a) demonstrates that the bonding
state in BC arrangement remains a localized state of the junction.
The DOS projected onto the Hydrogen atom
shows that this state is essentially unaffected by the proximity
of the electrodes. It indeed remains as a sharp peak located at
almost the same energy as for the isolated molecule, somewhat below
the lower Pt valence band edge. These results are confirmed by
also plotting the DOS projected onto the neighboring Pt atoms,
where the weight of such bonding state has decreased by
a factor of four. In contrast, the anti-bonding state is completely
spread, coupling strongly to the leads. We are
therefore led to conclude that the
bonding channel of conduction is completely closed down, having zero
transmission, while the antibonding state must be open. Moreover,
a closer examination of the Density of States projected onto single
orbitals and also of the Hamiltonian shows
that such antibonding state hybridizes only with the $s$ and
$d_{z^2-r^2}$ orbitals of the leads, as expected by symmetry.
Indeed, we find that each Hydrogen atom gains 0.27 electrons, that
populate the anti-bonding state DOS below the Fermi energy.

\begin{figure}
\includegraphics[width=\columnwidth,height=6cm]{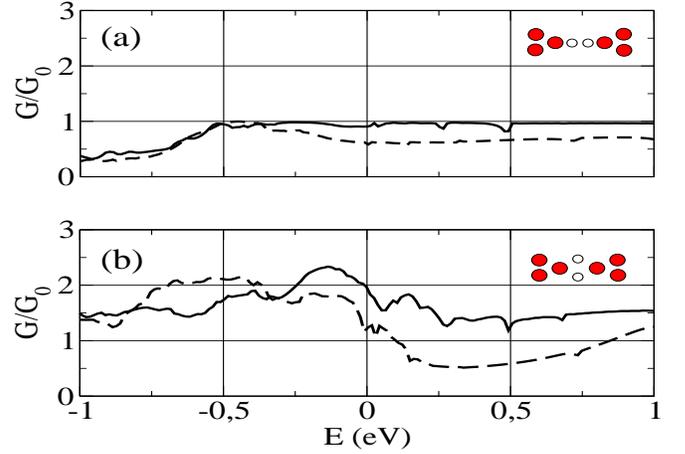}
\caption{\label{Fig4} Transmission coefficients in (a) BC, and (b)
PC settings at their equilibrium distances. Solid and dashed lines
indicate platinum and palladium, respectively. Energies are all referred to E$_F$.}
\end{figure}

The situation of the PC arrangement is somehow different. In this
case the H$_2$ molecule again hybridizes both with the $s$ and
$d_{z^2-r^2}$ orbitals of the tip of the electrodes. However, Fig.
3(b) shows that the bonding state of the molecule now participates
in the chemical bond to the leads. First, notice that the
corresponding peak in the DOS has slightly moved up in energies
and splitted, to place itself at the band edge of the d-band of
the Pt electrodes. Second, the amplitude of the peak does not
decrease when moving from the Hydrogen to the Pt atom at the apex.
This fact demonstrates that the bonding state becomes delocalized
and therefore opens up as a conduction channel. These features
will become more apparent for Pd electrodes, as we shall discuss
below.

Let us now turn our attention to the electron transmission through
these states in the BC configuration. We have shown above that the
bonding channel is completely closed down. That means that only
the anti-bonding channel can swap electrons between both
electrodes and, therefore, the conductance can be as large as 1.0
$G_0$ at most. We find numerically that the conductance is almost
flat, and approximately equal to $0.9\, G_0$, for a wide range of
energies around E$_F$ within the Pt bandwidth, see Fig. 4.
Moreover, we have found that T(E) remains essentially the same for
the the whole range of distances where BC is stable until the
contact breaks.  This almost perfect and constant transmission of the
anti-bonding channel leads to the prediction of a narrow peak in
conductance histograms, with low fluctuations.

In contrast, for the PC arrangement we obtain a conductance of
about $2\,G_{0}$. In this case $T(E)$ shows strong variations around E$_F$, and
achieves values even higher than 2 for distances around
$d_{eq}\approx 9.5$ \AA~ which implies the existence of more than
two channels available for conduction. This arises from direct
transport through Pt atoms at the apices of the pyramids, which at
such a short distance have a small but finite overlap. Therefore
if the PC is realized experimentally, we expect the conductance
histograms to display a peak centered at 2~$G_0$, with rather
strong fluctuations around that value. This result is in
contradiction with that of Garc\'{\i}a~\cite{Gar04}.

We now discuss the stability of the equilibrium configuration of
the junction. Our simulations confirm that the PC has a lower
energy than the BC as shown in Fig. 5, in agreement with other
authors \cite{Gar04,Thy04}. At first sight, this seems to be in
contradiction with the experimental data, since the PC
configuration has a conductance of about 2~$G_0$ instead on
1~$G_0$. However, a closer look into the energetics of the problem
reveals two important features.

\begin{figure}
\includegraphics[width=\columnwidth,height=6cm]{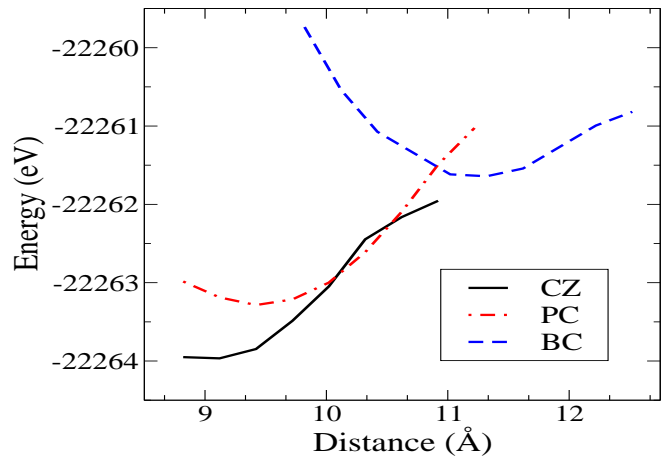}
\caption{\label{Fig5} Energy-cohesion curves $E(d)$: energy $E$ of different atomic
configurations of the junction as a function of distance $d$. Solid,
dashed and dash-dotted lines indicate CZ, BC and PC arrangements.}
\end{figure}

First, the last Pt atoms of the tips may prefer to bond in a
zigzag configuration, thereby continuing the fcc alignment. We
have performed further simulations using such zigzag arrangements
and found that their energy is substantially lower than that of
their aligned counterparts whenever the two Pt atoms at the tip
bond directly. In contrast, the energy-cohesion curves are
insensitive to the alignment of the pyramids if a Hydrogen
molecule sits between the two Pt atoms, independently of whether
we choose BC or PC configurations.

Secondly, if the distance between electrodes is too short, the
Hydrogen molecule may prefer to gain energy by bonding elsewhere
on the Pt surface. We have therefore simulated zigzag
configurations (CZ) where the Hydrogen molecule is on the surface
of one of the leads, as shown in Fig. 2 (c). We have then relaxed
coordinates to let both atoms achieve their equilibrium position
on top of the triangular sites between the pyramid and the lead.
Figure 5 shows that the CZ arrangement is indeed lowest in energy
for short distances, as expected, while there is a window of about
1 \AA~, where the PC is stable and competes in energy with all the
different possible realizations of CZ. We find that CZ snaps for
distances larger than 11 \AA$\,$ (and gives zero conductance). For
these distances the BC becomes then the most stable configuration.

We are therefore able to explain why the conductance histograms of
the break junctions do not show a peak at 2 $G_0$, since the PC is
unstable at short distances against configurations where the H$_2$
molecule is absorbed on the surface of the electrodes, and against the BC
configuration at large distances, where all other arrangements break.
Therefore we conclude that the typical experimental situation is
that of a H$_2$ molecule bridging the Pt leads in a BC
configuration, with a conductance of approximately $G_0$.

\begin{figure}
\includegraphics[width=7cm,height=7cm]{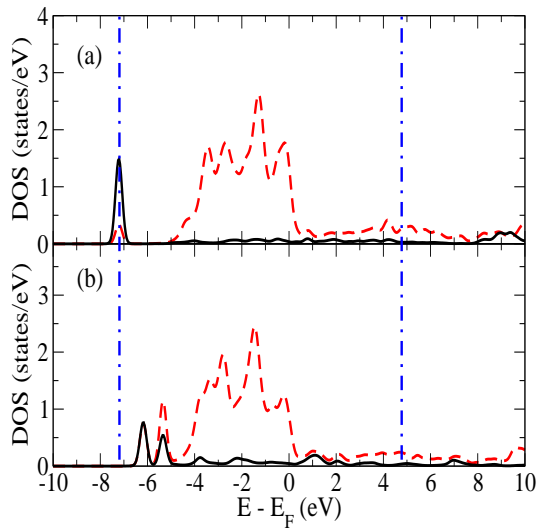}
\caption{\label{Fig6} Pd electrodes: DOS projected on one of the
Hydrogen atoms and the Pd atom at one of the two apexes (solid and
dashed lines, respectively) in (a) BC and (b) PC configurations.
Dashed-dotted vertical lines: position of the bonding and
anti-bonding energy levels of an isolated H$_2$ molecule. Energies
have been referred to the Fermi energy of the Pd leads.}
\end{figure}

Based on the above results, we propose the following evolution of
a Pt break junction in presence of Hydrogen as a function of the
Pt-Pt lead separation. For short distances (up to about 10 \AA~)
the junction arranges itself as a zigzag Pt point contact. H$_2$
molecules attach to the Pt surface at the triangular sites,
away from the atomic constriction. In this case,
electrons travel between leads through several Pt channels and the
conductance is much larger than $G_0$. As the separation between the
contacts increases from 10.0 to 11.0 \AA, the PC configuration
may take over in a few of the pulling cycles of the break junction. For
distances larger than about 11 \AA~, all configurations  other than
BC snap. The junction therefore arranges in a BC configuration
before eventually also snapping at a distance 12.5 \AA. This latter
arrangement provides a single channel conductance of about
0.9~$G_0$, that results in the sharp peak found in the
experimental conductance histograms \cite{Smi02}.

Finally we consider the case of palladium leads, whose bonding
properties are similar to those of Pt. For the BC, we find again
that the bonding state of the H$_2$ molecule does not participate
in the chemical bond. Fig. 6(a) shows that the bonding state
remains a sharp peak that is now clearly dettached from the lower
edge of the d-band of Pd. The amplitude of the peak in the Density
of States again decays very fast as one moves away from the
Hydrogen atom. We find as before that the anti-bonding state
strongly hybridizes with the $s$ and d$_{z^2-r^2}$ orbitals of
neighboring Pd atoms, so that electronic conduction is carried
again by the anti-bonding channel. For the PC configuration, the
bonding state hybridizes, leading to the opening of a second
conduction channel. Fig. 6(b) shows that the peak of the bonding
state moves up in energies by 2 eV when the molecule hybridizes
with the Pd electrodes in the PC configuration, placing itself at
the band edge of the d-band. Moreover, the amplitude of the peak
is almost the same for Hydrogen and for the Pd atom at the apex of
the electrode.

The transmission coefficients for the BC and PC are presented as
dashed lines in Fig. \ref{Fig4}. In complete agreement with the
experimental data \cite{Cso04}, we find a conductance of
0.6~$G_{0}$ for the BC arrangement at distances around
$d_{eq}\approx 11$ \AA\.. This means that the
transmission through the anti-bonding channel is not so perfect as
in the case of Pt electrodes. We also find that the conductance of
the PC is somewhat larger than 1 $G_0$ for distances around
$d_{eq}\approx 9.2$. Interestingly we obtain that
in the BC configuration, Pd atoms in the tip acquire a net
magnetic moment for distances larger than about $12$ \AA. This
results in the spin polarization of the current, since the
transmission coefficients are different for up and down spins.
Experimental evidence for spin-polarized transport in Pd point
contacts has been recently provided \cite{ugarte}.

The different transmissions found for the isostructural Pt and Pd
leads may be understood in terms of the alignment of the centers
of the s-d band of the leads and the hybridized anti-bonding
state. Indeed, a simple one-dimensional tight-binding model
consisting of two leads connected to a single level \cite{Car71}
shows that the transmission coefficient $T(E)$ has a long plateau
with value close to 1, if the on-site energies of the level and
the atoms at the leads differ only slightly. The plateau is
rounded and the transmission decreases as the difference between
both energies is made larger. This behavior is evident  in Fig. 4.

In conclusion, we find that  H$_2$ molecules typically bind to
Pt or Pd leads in the BC configuration. The bonding state of the
H$_2$ molecule does not participate in the chemical
bond, and does not contribute to the low bias conductance.
The transmission coefficient is close to 1 $G_0$ for Pt leads,
since the energy of the hybridized anti-bonding orbital is close to
the center of the $s$-$d$ band of Pt. Such alignment
is not realized for Pd electrodes,
leading to a smaller conductance.

\begin{acknowledgments}
VMGS thanks the Spanish Ministerio Espa\~nol de
Educaci\'on, Cultura y Deporte for the fellowship AP2000-4454.
ARR acknowledges a fellowship from Enterprise Ireland
(grant EI-SC/2002/10).
Financial support from Spanish Ministerio de Educaci� y Ciencia
(grant BFM2003-03156), UK EPSRC, Irish SFI (grant 02/IN1/I175) and the EU
network MRTN-CT-2003-504574 are also acknowledged. Traveling has been
sponsored by the Royal Irish Academy under the International
Exchanges Grant scheme.
\end{acknowledgments}

%\bibliography{PtPdH2}

\begin{thebibliography}{99}
\bibitem{Basis} SIESTA replaces core electrons by pseudopotentials and describes
valence electrons by an atomic-like basis set. We use here a
double-$\zeta$ basis for Hydrogen and double-$\zeta$ polarized
bases for platinum and palladium. The exchange and correlation
potential is calculated using the local density approximation
(LDA).
\bibitem{k-points} Tests performed with 12 k-points provided essentially the same results.
4 k-points are enough to avoid spurious peaks or gaps in the transmission coefficients.
\end{thebibliography}

\end{document}